\documentclass[iicol, sn-mathphys, Numbered]{sn-jnl}% Math and Physical Sciences Reference Style

\usepackage{graphicx}%
\usepackage{array, multirow}%
\usepackage{hhline}
\usepackage{tabularray}
\usepackage{amsmath,amssymb,amsfonts}%
\usepackage{amsthm}%
\usepackage{mathrsfs}%
\usepackage[title]{appendix}%
\usepackage{xcolor}%
\usepackage{textcomp}%
\usepackage{manyfoot}%
\usepackage{booktabs}%
\usepackage{algorithm}%
\usepackage{algorithmicx}%
\usepackage{algpseudocode}%
\usepackage{listings}%
\usepackage{graphicx}
\usepackage{titlesec}
\usepackage{caption}
\usepackage{nicefrac}
\usepackage{makecell}
\usepackage[T1]{fontenc}

\titleformat{\section}
  {\normalfont\Large\bfseries}{\thesection}{1em}{}[{\titlerule[0.8pt]}]

\theoremstyle{thmstyleone}

\begin{document}

%%==================================%%
%%              title               %%
%%==================================%%

\title[Article Title]{A deep-learning model for one-shot transcranial ultrasound simulation and phase aberration correction}

%%==================================%%
%%             authors              %%
%%==================================%%


\author*[1,]{\fnm{Kasra} \sur{Naftchi-Ardebili}}\email{knaftchi@stanford.edu}
\equalcont{These authors contributed equally to this work.}

\author*[2,]{\fnm{Karanpartap} \sur{Singh}}\email{karanps@stanford.edu}
\equalcont{These authors contributed equally to this work.}

\author[3,4]{\fnm{Gerald} \sur{R. Popelka}}\email{gpopelka@stanford.edu}

\author[1,2,3]{\fnm{Kim} \sur{Butts Pauly}}\email{kimbutts@stanford.edu}

\affil[1]{\orgdiv{Department of Bioengineering}, \orgname{Stanford University}, \orgaddress{\city{Stanford, CA}, \postcode{94301}}}

\affil[2]{\orgdiv{Department of Electrical Engineering}, \orgname{Stanford University}, \orgaddress{\city{Stanford, CA}, \postcode{94301}}}

\affil[3]{\orgdiv{Department of Radiology}, \orgname{Stanford University}, \orgaddress{\city{Stanford, CA}, \postcode{94301}}}

\affil[4]{\orgdiv{Department of Otolaryngology}, \orgname{Stanford University}, \orgaddress{\city{Stanford, CA}, \postcode{94301}}}

%%==================================%%
%%       unstructured abstract      %%
%%==================================%%

\abstract{Transcranial ultrasound (TUS) has emerged as a promising tool in both clinical and research settings due to its potential to modulate neuronal activity, elicit blood brain barrier opening, facilitate targeted drug delivery via nanoparticles, and perform thermal ablation, all non-invasively. The method delivers focused ultrasound waves to precise regions at any location in the brain, enabling targeted energy deposition. The medical importance of transcranial ultrasound is evidenced by the upwards of fifty ongoing clinical trials, covering conditions such as opioid addiction, Alzheimer's disease, dementia, epilepsy, and glioblastoma. In addition to careful design of ultrasound parameters, treatments with transcranial ultrasound require precise computation of the location and pressure at the focal spot. However, the heterogeneity of the skull alters the incident ultrasound beams and if uncorrected, poses the risk of off-target sonication or inadequate energy delivery to the neural tissue. For clinical settings, this phase aberration correction must be done within a few seconds. However, physics-informed simulation software suffer from an inherent trade-off between accuracy and efficiency. As such, commercial devices currently rely on faster methods that provide the short runtime necessary for clinical use but may not always achieve the high level of accuracy needed for optimal TUS delivery. We present TUSNet, the first end-to-end deep learning approach to solve for the pressure field as well as phase aberration corrections, without being bound to this inherent trade-off between accuracy and efficiency. TUSNet can compute the 2D transcranial ultrasound pressure field and phase corrections within 21 milliseconds (over 1200$\times$ faster than k-Wave, a MATLAB-based acoustic simulation package), while achieving $98.3\%$ accuracy in estimating the absolute value of the peak pressure at the focal spot with a mean positioning error of only $0.18\ \mathrm{mm}$, when compared to a ground truth from k-Wave.}

%%================================%%

\maketitle

%%==================================%%
%%           introduction           %%
%%==================================%%

\clearpage % start on next page
\section*{Main}

In recent years, transcranial ultrasound has been utilized for research and clinical applications, owing to its non-invasive approach to neural tissue modulation, blood brain barrier opening, drug delivery via nanoparticles, and thermal ablation. With its roots dating back to the pioneering work of Fry et al. \cite{Fry1959IntenseSystem}, transcranial ultrasound has since been a focal point of interest within the research community, leading to many key advancements in the field and allowing safe and effective treatment in humans \cite{Legon2018NeuromodulationThalamus, Zhang2021TranscranialAnimals, Lee2016TranscranialCortex, Ghanouni2015TranscranialApplications, Choi2019TherapeuticDiseases, Leinenga2016UltrasoundApplications, Piper2016FocusedReview, Elias2016ATremor, Fishman2017FocusedDisease, Lipsman2018BloodbrainUltrasound, Lipsman2013MR-guidedStudy, Na2015UnilateralDisease, Jeanmonod2012TranscranialPain, Bond2017SafetyDisease, Martinez-Fernandez2018FocusedStudy, Mueller2014TranscranialDynamics, Tufail2011UltrasonicUltrasound, Legon2014TranscranialHumans}. Moreover, at the moment, there are upwards of fifty ongoing clinical trials to assess the effectiveness of transcranial ultrasound in noninvasive treatment of a range of disorders including epilepsy, opioid addiction, Alzheimer's disease, glioblastoma, attention deficit hyperactivity disorder, and obsessive compulsive disorder \cite{Studiesoftranscranialultrasound.Retrievedfrom2024ClinicalTrials.gov}. Despite these significant strides, the optimization of transcranial ultrasound parameters for individualized treatments remains a significant challenge due to the complexity of ultrasound propagation in the human skull \cite{Fink1992TimePrinciples, Kyriakou2014AUltrasound., Leung2019AUltrasound}.

The successful application of transcranial ultrasound relies heavily on the precise computation of the location and pressure at the focal spot. Ideal precision in TUS is on the sub-millimeter order, and depends on the target size, intended outcome, and transducer frequency, but the obtained precision in focal positioning and energy deposition is greatly influenced by the skull's heterogeneity \cite{, Blackmore2019UltrasoundSafety, Mueller2017NumericalUltrasound, Pinton2011AttenuationBone}. The skull introduces phase aberrations to the ultrasound wavefront, which affect the location and intensity of the ultrasound focus within the brain, potentially leading to off-target effects or reduced efficacy. Physics-based simulation methods, such as the $k$-space pseudospectral method \cite{Treeby2012ModelingMethod} and hybrid angular spectrum (HAS) method \cite{Vyas2008UltrasoundMethod, Leung2021TranscranialMethod}, provide powerful tools to address this challenge by modeling the acoustic propagation of the ultrasound waves through the skull. Notably, for clinical relevance, the accurate computation of the location and pressure at the focal spot must be performed within a few seconds. However, these physics-informed simulations are computationally intensive and can take minutes to hours to converge, hindering their use in real-time clinical settings.

Currently, commercial devices often rely on faster methods like ray tracing to predict the focus location and pressure. While these methods are computationally efficient, they may not always achieve the high level of accuracy needed for optimal TUS delivery \cite{Leung2021TranscranialMethod}. Leung et al. showed that while the InSightec Exablate 4000 was capable of computing the phase corrections for all of its 1,024 elements in approximately 2 seconds, it only recovered $71 \pm 15\%$ of the pressure at the target (compared against hydrophone ground truth measurements), while incurring a positioning error of $0.72 \pm 0.47\ \mathrm{mm}$. On the other hand, HAS recovered $86 \pm 5\%$ of the pressure at the target, and at the same time reduced the positioning error down to $0.35 \pm 0.09\ \mathrm{mm}$. But this significant improvement in targeting efficacy and accuracy came at a cost: what took InSightec's proprietary ray tracing 2 seconds to compute, took HAS 30 minutes \cite{Leung2021TranscranialMethod}. With the suboptimal positioning accuracies achievable with fast but less accurate  methods such as ray tracing, clinicians are forced to iteratively adjust the position of the transducer on the patient until they see the intended focal spot using imaging methods such as Magnetic Resonance Thermometry \cite{Marx2017SpecializedBrain, Rieke2008MRThermometry}. The ideal, which remains unresolved to this date, is to reach the accuracies achieved by HAS \cite{Leung2021TranscranialMethod}, but with run times similar to that of ray tracing.

Therefore, a significant gap remains between the need for rapid and accurate predictions of the transcranial ultrasound pressure field and the currently available computational methods. While machine learning is poised to address this inherent trade-off between accuracy and efficiency, few such attempts have been made to date. Shin et al. \cite{Shin2023Multivariable-IncorporatingSimulation} developed a super-resolution neural network that transformed $1$-mm low resolution voxels into $0.5$-mm higher resolution voxels, improving the simulation run time by a remarkable 86.91\%, but it did not correct for phase aberrations and utilized a single-element transducer. Choi et al. \cite{Choi2022DeepConcept} proposed an interesting model that identified the ideal $x,y,z$ location for a single-element transducer placement, given a desired pressure field. But this approach relied on prior knowledge of the desired transcranial pressure field, and the model was not tasked with computing and correcting for the phase aberrations. 

\begin{figure*}[ht!]
\centering
\includegraphics[width=1.0\textwidth]{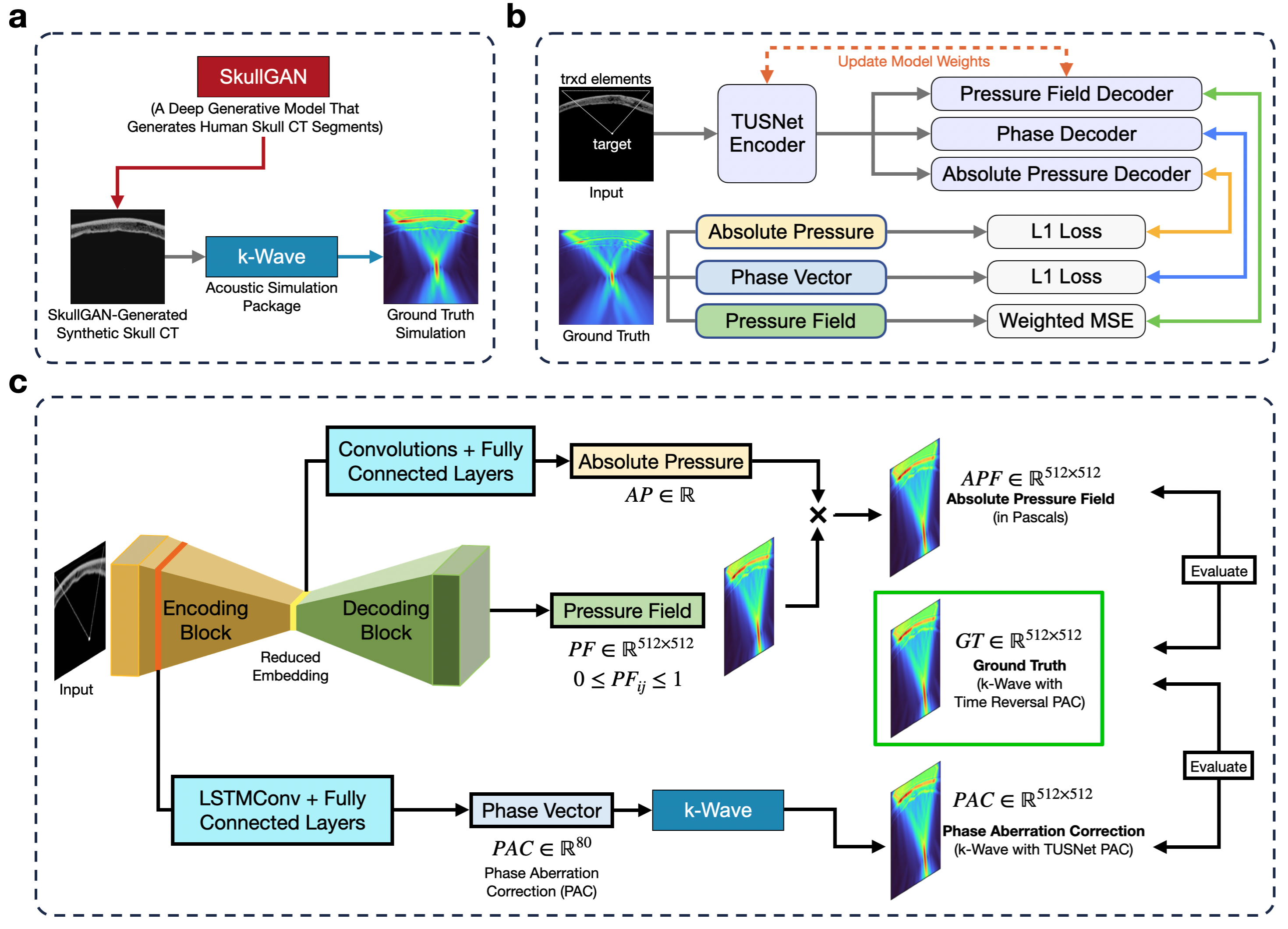}
\caption{\textbf{TUSNet training (a-b) and inference (c) pipelines.} \textbf{a}. TUSNet is trained entirely on synthetic skull CT images generated by SkullGAN, a generative model that, by learning the underlying data generating distribution of real human skull CTs, produces synthetic 2D skull CT segments. Transcranial ultrasound propagation is simulated through the skull using k-Wave to generate the ground truth training data. \textbf{b}. TUSNet comprises a multi-task encoder-decoder architecture, with three decoders individually predicting the ultrasound \emph{pressure field}, \emph{phase vector} (phase aberration corrections), and \emph{absolute pressure} (peak focal pressure), given an input. The input comprises the skull CT segment, target location, and locations of the transducer elements lying above the skull segment. We first trained the \emph{TUSNet Encoder} and \emph{Pressure Field Decoder} to predict the ultrasound pressure field, after which the \emph{TUSNet Encoder} was frozen and the \emph{Phase Decoder} and \emph{Absolute Pressure Decoder} were trained independently, each using a separate loss. \textbf{c}. At inference, the \emph{Absolute Pressure} ($\in \mathbb{R}$) was derived from the the \emph{Reduced Embedding}, and the normalized \emph{Pressure Field} from the \emph{Decoding Block}. These were multiplied to produce the \emph{Absolute Pressure Field} in Pascals. The \emph{Phase Vector} was estimated by applying \emph{LSTMConv} and fully connected layers to the \emph{Encoding Block}'s first output, and evaluated with k-Wave simulations.}
\label{intro}
\end{figure*}

\begin{figure*}[t]
\centering
\includegraphics[width=1.0\textwidth]{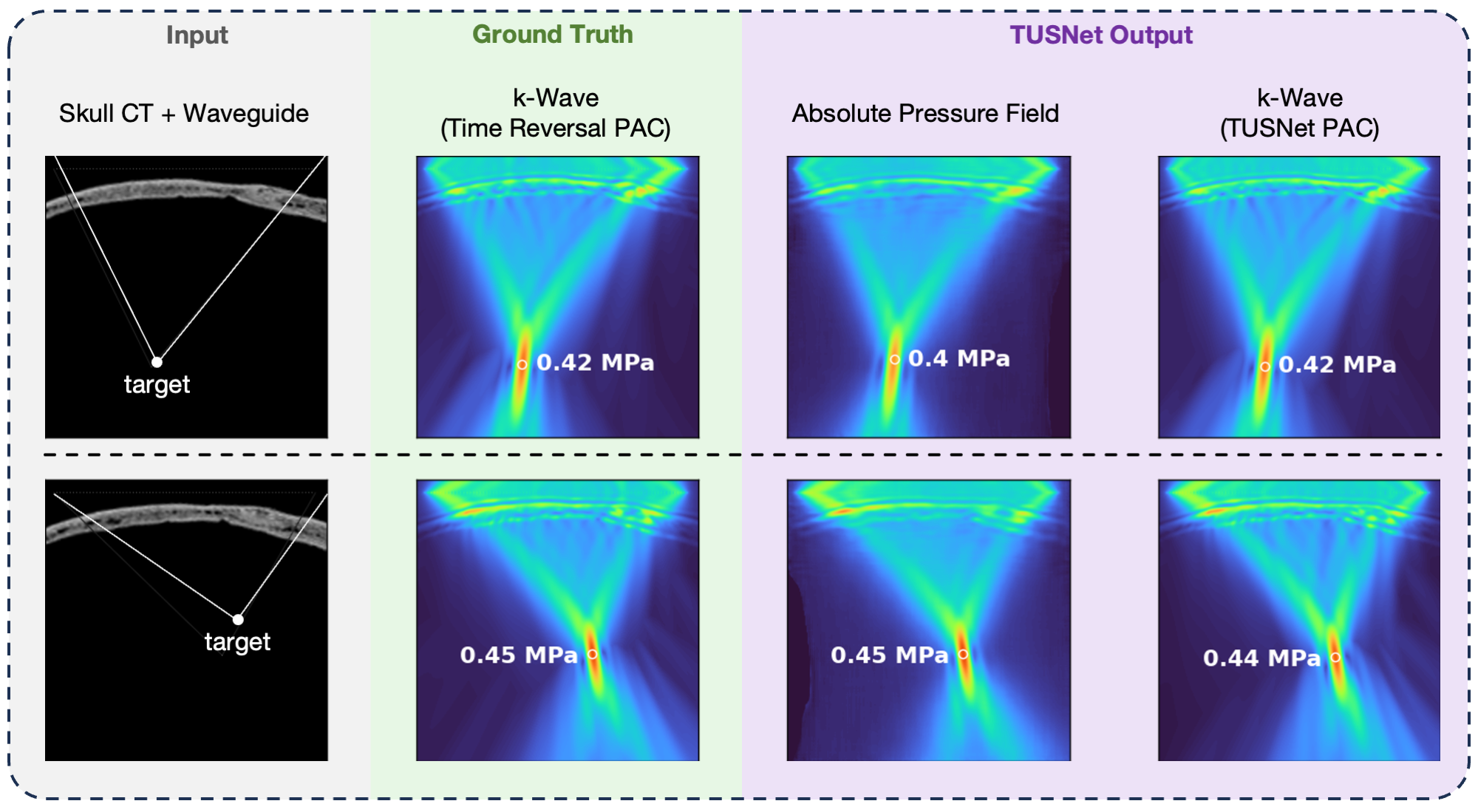}
\caption{\textbf{Comparison between ground truth and TUSNet outputs (columns) for two examples (rows).} \textbf{Input:} consists of the transducer elements above the skull (background removed), a waveguide, and the target. \textbf{Ground Truth:} k-Wave-simulated ground truth phase aberration-corrected pressure field using time reversal. \textbf{TUSNet Output for Absolute Pressure Field:} in Pascals, rather than normalized values. \textbf{TUSNet Output for Phase Vector:}  simulated by k-wave based on the TUSNet phase vector, rather than time reversal.}
\label{tusnet_outputs}
\end{figure*}

In this paper, we propose TUSNet (Fig \ref{intro}), a machine learning approach designed to bridge this gap between accuracy and efficiency. Leveraging the power of sequence-based deep learning architectures, TUSNet is the first end-to-end model capable of computing the 2D transcranial ultrasound pressure field and phase aberration corrections for an 80-element phased array transducer within 0.0207 seconds (one NVIDIA A4000 GPU \footnote{average of 10 runs, 0.062 seconds on a MacBook Pro w/ M1 Pro}) while maintaining the desired accuracy for focal positioning and pressure recovery. The TUSNet phase corrections on average recovered 98.3\% of the focal pressure (compared to ground truth data from k-Wave, a widely used acoustic simulation package), while maintaining a mean positioning error of $0.18\ \mathrm{mm}$. Although TUSNet is a proof of concept in showing the possibility of performing an end-to-end transcranial pressure field simulation and phase aberration correction in 2D, it provides the blueprint for its 3D extension. We believe that the introduction of TUSNet, in conjunction with the other proposed models so far \cite{Naftchi-Ardebili2023SkullGAN:Networks, Shin2023Multivariable-IncorporatingSimulation, Choi2022DeepConcept}, will revolutionize the field of TUS by paving the way for development of clinical models that enable real-time phase aberration correction and pressure field simulations with high accuracies, significantly streamlining patient treatments.

%%==================================%%
%%             results              %%
%%==================================%%

\begin{table*}[h]
\resizebox{\textwidth}{!}{%
\begin{tabular}{|c|cccccccc|}
\hline
\multirow{2}{*}{} &
  \multicolumn{8}{c|}{Absolute Pressure Field Performance on the Test Set} \\ \hhline{|~|--------|}    
 &
  \multicolumn{2}{c|}{Focal Area Error (\%)} &
  \multicolumn{2}{c|}{Pressure Error (\%)} &
  \multicolumn{4}{c|}{Focal Position Error (mm)} \\ \hline
Model Trained on: &
  \multicolumn{1}{c|}{Percent Error} &
  \multicolumn{1}{c|}{IoU} &
  \multicolumn{1}{c|}{Focal} &
  \multicolumn{1}{c|}{Peak} &
  \multicolumn{1}{c|}{Euclidean} &
  \multicolumn{1}{c|}{Hausdorff} &
  \multicolumn{1}{c|}{Axial} &
  Lateral \\ \hline
Synthetic CTs &
  \multicolumn{1}{c|}{8.27 $\pm$ 5.43} &
  \multicolumn{1}{c|}{86.95 $\pm$ 7.05} &
  \multicolumn{1}{c|}{6.07 $\pm$ 4.0} &
  \multicolumn{1}{c|}{5.87 $\pm$ 3.95} &
  \multicolumn{1}{c|}{0.30 $\pm$ 0.26} &
  \multicolumn{1}{c|}{0.15 $\pm$ 0.08} &
  \multicolumn{1}{c|}{0.28 $\pm$ 0.26} &
  0.06 $\pm$ 0.07 \\ \hline
\end{tabular}}
\caption{\textbf{TUSNet \emph{absolute pressure field} performance on a test set of real skull CTs.} The model was trained on 180,432 SkullGAN-generated synthetic CTs only. Conversely, the test set included real skull CT segments only that were obtained from three separate
patients never seen by either TUSNet or SkullGAN.}
\label{tusnet_metrics_table_1}
\end{table*}

\begin{table*}[h]
\resizebox{\textwidth}{!}{%
\begin{tabular}{|c|cccccccc|}
\hline
\multirow{2}{*}{} &
  \multicolumn{8}{c|}{Phase Aberration Correction Performance on the Test Set} \\ \hhline{|~|--------|}    
 &
  \multicolumn{2}{c|}{Focal Area Error (\%)} &
  \multicolumn{2}{c|}{Pressure Error (\%)} &
  \multicolumn{4}{c|}{Focal Position Error (mm)} \\ \hline
Model Trained on: &
  \multicolumn{1}{c|}{Percent Error} &
  \multicolumn{1}{c|}{IoU} &
  \multicolumn{1}{c|}{Focal} &
  \multicolumn{1}{c|}{Peak} &
  \multicolumn{1}{c|}{Euclidean} &
  \multicolumn{1}{c|}{Hausdorff} &
  \multicolumn{1}{c|}{Axial} &
  Lateral \\ \hline
Synthetic CTs &
  \multicolumn{1}{c|}{8.52 $\pm$ 4.61} &
  \multicolumn{1}{c|}{85.58 $\pm$ 5.85} &
  \multicolumn{1}{c|}{1.71 $\pm$ 1.09} &
  \multicolumn{1}{c|}{1.13 $\pm$ 0.68} &
  \multicolumn{1}{c|}{0.59 $\pm$ 0.35} &
  \multicolumn{1}{c|}{0.18 $\pm$ 0.07} &
  \multicolumn{1}{c|}{0.57 $\pm$ 0.35} &
  0.10 $\pm$ 0.10 \\ \hline
\end{tabular}}
\caption{\textbf{TUSNet \emph{phase aberration correction} performance on a test set of real skull CTs.} Phase vectors obtained with TUSNet were input to k-Wave as the element-wise phase aberration corrections. The forward simulation results of k-Wave were used to assess the accuracy of the TUSNet phase vectors against that of time reversal (ground truth). The test set comprised real skull CT segments only, and were obtained from three separate patients never seen by either TUSNet or SkullGAN.}
\label{tusnet_metrics_table_2}
\end{table*}

\section*{Results}

The TUSNet outputs, consisting of the normalized pressure field, peak absolute pressure, and phase aberration corrections, are mostly decoupled with varying degrees of parameter sharing between them. As described in Fig. \ref{intro}c, the normalized pressure field utilizes the entire encoding and decoding blocks; the absolute pressure re-scaling factor shares only the encoding and reduced embedding parameters with the normalized pressure field; and the phase vector arm of TUSNet shares only the first LSTM-Conv cell of the encoding block. The product of the normalized pressure field and the absolute pressure re-scaling factor yields the \emph{absolute pressure field} output of TUSNet. The phase vector output of TUSNet is simply an 80-dimensional vector, corresponding to the corrections applied to each of the 80 transducer elements. In order to be able to evaluate this phase vector with the same criteria and methods used for evaluating the \emph{absolute pressure field}, we implemented transcranial k-Wave simulations with the TUSNet phase vector, instead of the ground truth method of time reversal, and labeled this output as the \emph{phase aberration correction} output of TUSNet. As such, we present our results separately: for every metric, we evaluated the \emph{absolute pressure field} and \emph{phase aberration correction} of TUSNet independently.

\subsection*{Qualitative Accuracy}
Capturing the shape and location of the focal spot was not the sole objective in training TUSNet. The entirety of the wave pattern, including the side lobes as well as the hot spots inside the skull, were also replicated. Figure \ref{tusnet_outputs} showcases the quality of TUSNet's output by comparing them to the ground truth from the k-Wave simulation. Both the absolute pressure field and the phase aberration corrections estimated by TUSNet successfully replicated the shape and amplitude of the pressure field near the focus, while also capturing finer reflections in the skull and outside of the focal spot. 

\subsection*{Quantitative Accuracy}
\subsection*{I. Focal Area}
\subsubsection*{Percent Error}
The full width at half maximum (FWHM) of the absolute pressure field estimated by TUSNet deviated from that of the ground truth by 8.27\%. TUSNet's phase aberration correction had a similar performance at 8.52\%. These results are summarized in Fig \ref{tusnet_metrics_focal_area}a.

\subsubsection*{Intersection over Union (IoU)}
\begin{figure*}[t]
\centering
\includegraphics[width=1.0\textwidth]{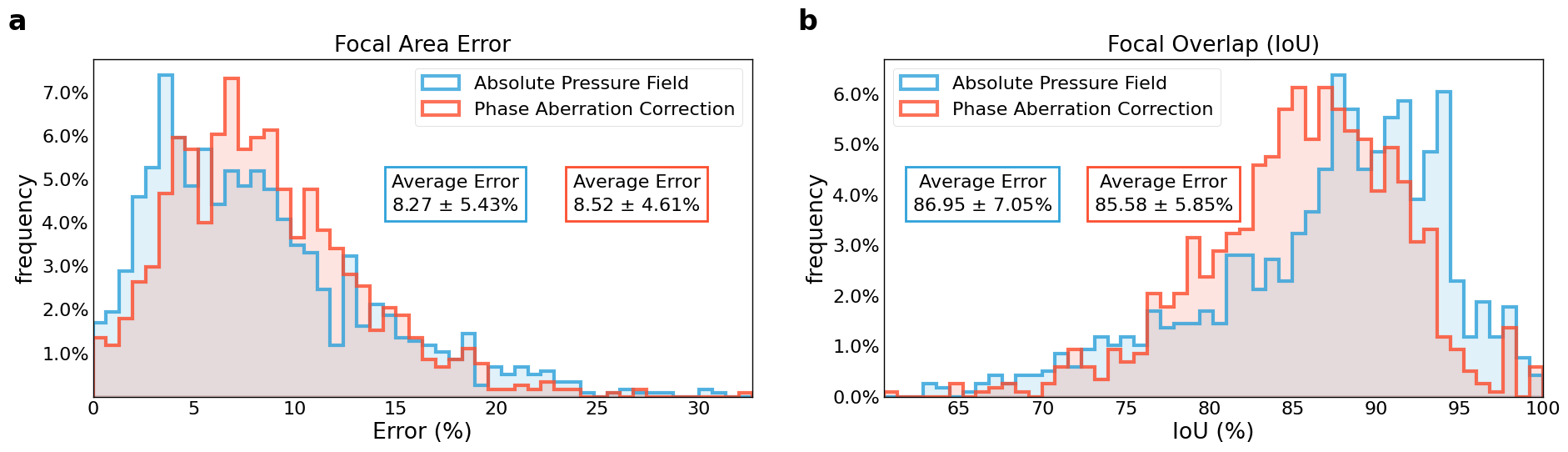}
\caption{\textbf{TUSNet performance for the absolute pressure field and phase aberration correction on estimating the focal area for 1,232 skull segments.} Focal areas were segmented out as ellipses at the FWHM of the pressure. \textbf{a}. Frequency distributions (mean and standard deviation values in boxes) of focal area error, evaluated by comparing the areas of the segmented focal spots. \textbf{b}. Frequency distributions (mean and standard deviation values in boxes) of focal overlap, evaluated by intersection over union for the segmented focal spots.}
\label{tusnet_metrics_focal_area}
\end{figure*}

On average, the TUSNet absolute pressure field had an 86.95\% IoU. The phase aberration correction component of TUSNet did not perform too differently, with a mean IoU of 85.58\% (Fig \ref{tusnet_metrics_focal_area}b). Previous work on using deep learning to estimate transducer location and orientation for a given, binarized, target ellipsoid, reported 74.49\% in IoU \cite{Choi2022DeepConcept}. Another earlier work employing a super-resolution neural network to convert low resolution ($1\ \mathrm{mm}$) TUS pressure fields to high resolution ($0.5\ \mathrm{mm}$), reported an 80.87\% IoU \cite{Shin2023Multivariable-IncorporatingSimulation}. These studies were in 3D, but neither of them corrected for phase aberrations and used only single-element transducers.

\begin{figure*}[t]
\centering
\includegraphics[width=1.0\textwidth]{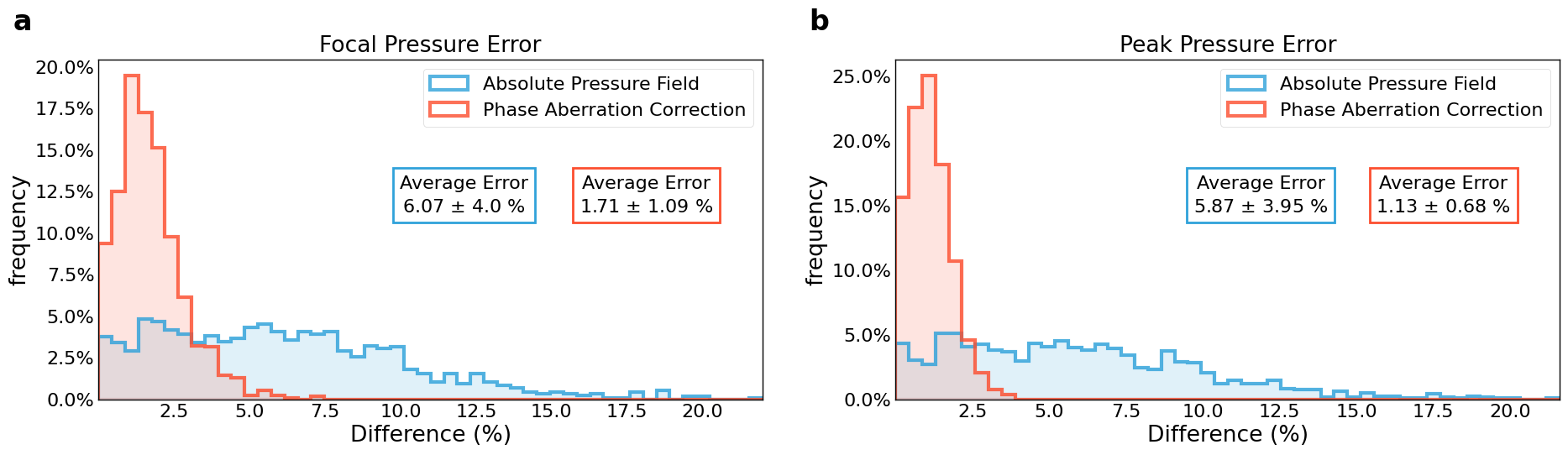}
\caption{\textbf{TUSNet performance for the absolute pressure field and phase aberration correction on estimating the focal and peak pressures for 1,232 skull segments.} \textbf{a}. Frequency distributions (mean and standard deviation values in boxes) of the focal pressure error, evaluated by comparing the pressure values at the foci. \textbf{b}. Frequency distributions of the peak pressure error (mean and standard deviation values in boxes), evaluated by comparing the peak pressures anywhere in the simulation frame.}
\label{tusnet_metrics_pressure}
\end{figure*}

\begin{figure*}[t]
\centering
\includegraphics[width=1.0\textwidth]{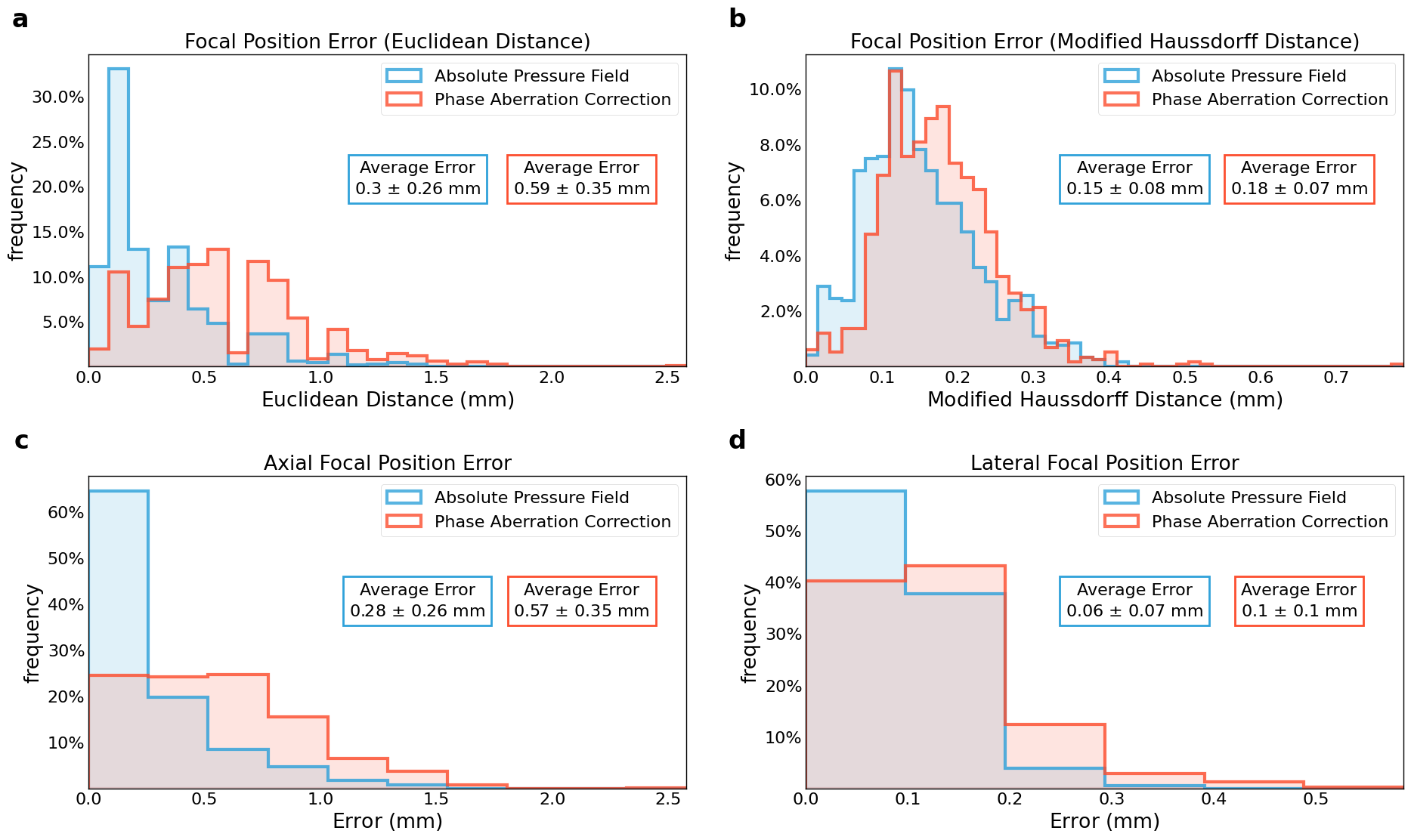}
\caption{\textbf{TUSNet performance for the absolute pressure field and phase aberration correction on estimating the focal position for 1,232 skull segments. Frequency distributions (mean and standard deviation values in boxes) of }\textbf{a}. Focal position error, evaluated by comparing the Euclidean distance between the foci. \textbf{b}. Focal position error, evaluated by the modified Hausdorff distance between the segmented ellipses at FWHM of the foci. \textbf{c}. Axial focal position error, determined by measuring the distance along the major axis of the FWHM ellipsoid. \textbf{d}. Lateral focal position, determined by measuring the error along the minor axis of the FWHM ellipsoid.}
\label{tusnet_metrics_position}
\end{figure*}

\subsection*{II. Pressure}
\subsubsection*{Focal Pressure Error}
The absolute pressure field estimated by TUSNet had a mean error of 6.07\% at the focus, compared to the numerical value of the ground truth pressure simulated by k-Wave. The phase aberration correction of TUSNet incurred only a 1.71\% average error (Fig \ref{tusnet_metrics_pressure}a). To our knowledge, this remarkable 98.3\% accuracy in recovering the pressure at the focus has no parallel in our field, as other deep learning attempts have only trained on normalized pressure values. HAS recovered 86\% of the pressure at the target \cite{Leung2021TranscranialMethod}, but it was a numerical simulation algorithm, not an end-to-end deep learning model. As such, it would still grapple with the accuracy-efficiency tradeoff mentioned earlier if executed at the $0.117\ \mathrm{mm}$ grid resolution of TUSNet. 

\subsubsection*{Peak Pressure Error}
Similar to the focal pressure error, the phase aberration correction accuracy of TUSNet outperformed its absolute pressure field in terms of peak pressure error, with a mean error of 1.13\% compared to 5.87\% (Fig \ref{tusnet_metrics_pressure}b).

\subsection*{III. Focal Position}
\subsubsection*{Euclidean Distance}
Computing the Euclidean distance between the single point with the peak pressure in TUSNet's outputs and k-Wave ground truth simulations resulted in a mean distance of $0.3\ \mathrm{mm}$ for the absolute pressure field, and $0.59\ \mathrm{mm}$ for the phase aberration correction (Fig \ref{tusnet_metrics_position}a). Choi et al. reported a $0.96\ \mathrm{mm}$ transcranial positioning error \cite{Choi2022DeepConcept}. We note that Euclidean distance doesn't take into account the location and contour error of the entire focal spot, unlike the Modified Hausdorff Distance.

\subsubsection*{Modified Hausdorff Distance (MHD)}
TUSNet's absolute pressure field was able to achieve a mean of $0.15\ \mathrm{mm}$ in focal positioning error, as determined with the MHD \cite{DubuissonAMatching}. The phase aberration correction had a similar accuracy, scoring an average MHD of $0.18\ \mathrm{mm}$. These results are shown in Fig \ref{tusnet_metrics_position}b. 

\subsubsection*{Axial Distance}
As shown in Fig \ref{tusnet_metrics_position}c, TUSNet's focal positioning errors were larger in the axial direction compared to the lateral direction, both in the absolute pressure field and phase aberration correction. The former had a mean axial error of $0.28\ \mathrm{mm}$, whereas the latter had a mean axial error of $0.57\ \mathrm{mm}$. 

\begin{figure*}[t]
\centering
\includegraphics[width=1.0\textwidth]{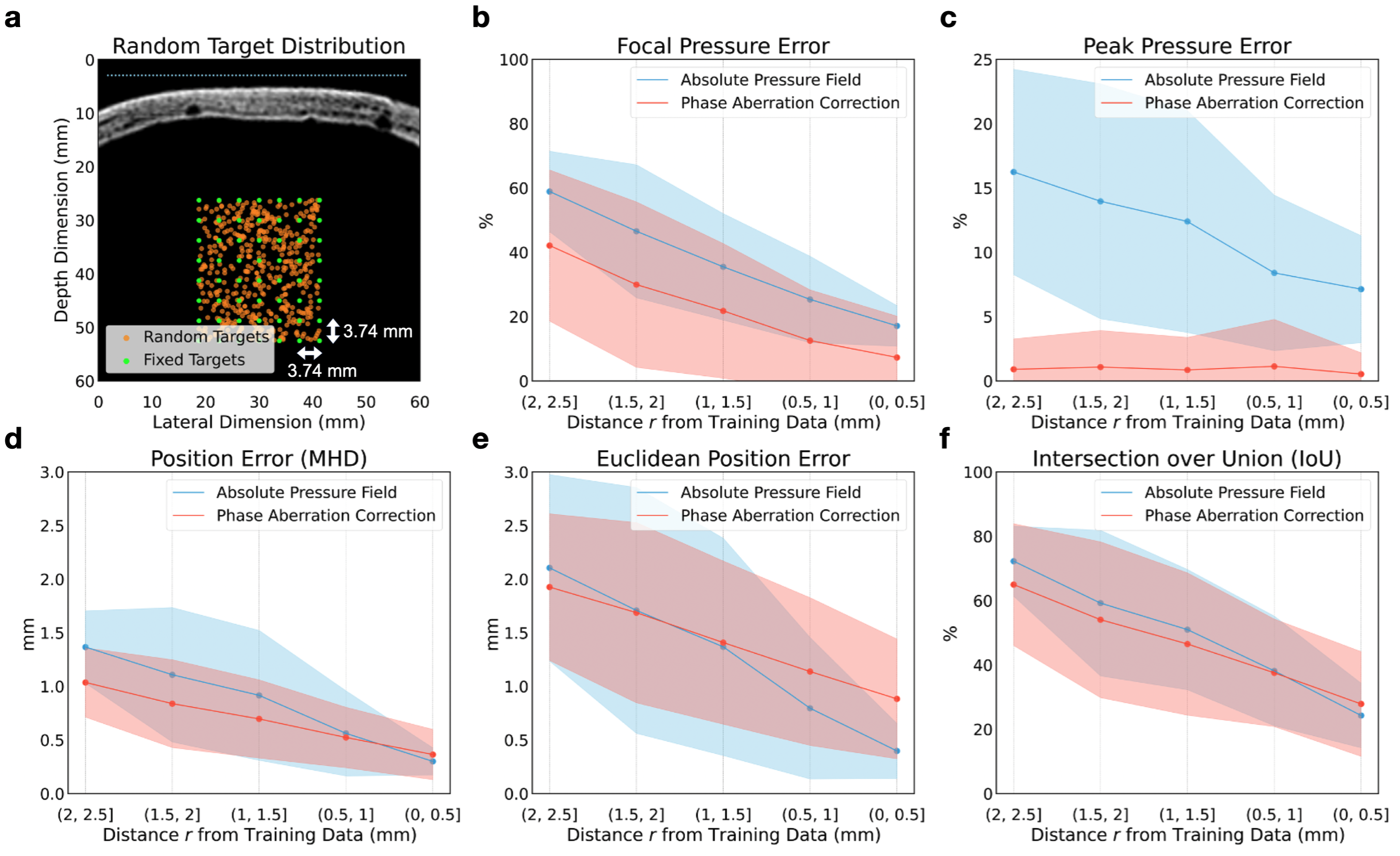}
\caption{\textbf{TUSNet performance for the absolute pressure field and phase aberration correction on 500 unseen random test targets}. \textbf{a.} The distribution of fixed training targets (green dots) as well as the distribution of the random test target points (orange dots). The blue dots on top of the real skull CT segment, taken from the test set, are the transducer elements. The 500 random test targets in orange are evaluated across 10 different skull segments, 50 targets per skull. Here we show only one of those real skull CT segments, but with all of the 500 test targets. \textbf{b-f.} Error (\% or mm)
as a function of radial distance of these random targets from the fixed training targets. These distances are in the form of disks, with their inner and out diameters specified along the $x$-axis.
}
\label{tusnet_random_targets}
\end{figure*}

\subsubsection*{Lateral Distance}

Compared to the axial error, the lateral positioning errors of TUSNet were much smaller. The absolute pressure field of TUSNet had a mean lateral error of $0.06\ \mathrm{mm}$, while its phase aberration correction had a mean lateral error of $0.1\ \mathrm{mm}$ (Fig \ref{tusnet_metrics_position}d). 

This stark difference between the axial and the lateral errors was primarily due to the fact that the axial beam profile is long and, within the tolerance of the model's loss function, several on-axis points immediately surrounding the focus also qualify as the peak pressure value. Conversely, because the lateral beam profile is much narrower in comparison to the axial beam profile, there is not much room for deviation from the true target location while maintaining a low loss value. 

\subsection*{IV. Error Analysis of Unseen Targets}

In training this version of TUSNet, our emphasis was on generalizing the model to unseen and varied skull CT segments. Therefore, the only variable in this setup was the skull CT - the target locations were held fixed. As shown in Fig. \ref{tusnet_random_targets}a, the training fixed targets were placed at a distance of $3.74\ \mathrm{mm}$ from one another, a spacing that is rather coarse for being able to generalize to unseen random targets in between. However, if we chose to place these targets closer to each other, we would not gain an understanding of the model's beam steering capabilities, as the 56 targets would be highly localized to one area. Conversely, if we wanted to keep the bounding box of the targets the same, and increase the number of target points inside it, we would fall short of the compute requirements for training on so many data points. This tradeoff would be alleviated with access to more computational resources. 

To analyze TUSNet's performance over randomly selected targets that did not coincide with the training points, as a function of radial distance away from the training points, we tested 50 randomized targets for each of 10 real skull CT segments. The objective was to see whether there was an inverse correlation between error and distance. In Fig. \ref{tusnet_random_targets}(b-f), we show that as the distance between the random targets and the fixed training targets reduces, errors also go down. This evidence confirms our assertion that if TUSNet is trained on random targets closely spaced from one another (under $0.5$ mm), it will generalize to both unseen targets and unseen skull CT segments. 

%%==================================%%
%%            discussion            %%
%%==================================%%

\section*{Discussion}

Transcranial ultrasound has emerged as a promising non-invasive technique with the potential to revolutionize neurotherapeutic interventions across a spectrum of neurological disorders. However, the journey towards realizing transcranial ultrasound's full clinical potential has been tempered by several challenges, one being the accurate and at the same time efficient modeling of ultrasound propagation through the complex anatomy of the human skull. 

In this study, we presented TUSNet, the first end-to-end deep-learning based approach to phase aberration correction and pressure field simulation, demonstrating near real-time results without compromising on precision. As a potential next step forward for transcranial ultrasound simulations, TUSNet provides a promising framework for allowing real-time, computationally-efficient, and patient-specific treatment planning. 

One main reason that such an end-to-end machine learning framework had not been applied to this problem is the stark absence of open training data of human skull CTs. We addressed this issue with SkullGAN \cite{Naftchi-Ardebili2023SkullGAN:Networks}, a deep-learning-based method for generating large quantites of synthetic skull CTs. Along with providing a sufficiently large dataset to train TUSNet, synthetic data effectively eliminates concerns about patient privacy, as no real patient data is used for training.

As the first model of its kind, TUSNet signifies a considerable leap forward towards clinically viable real-time transcranial ultrasound simulations. A logical next step is transitioning TUSNet to a 3D model. This would involve multiple layers of complexity: there would be an increased demand for very large datasets consisting of whole skull CT scans, corresponding high-resolution acoustic simulations, and inevitably, a larger neural network capable of handling 3D data. The first of these challenges could be facilitated with synthetic training data, similar to that used in this study. However, a 3D variant of SkullGAN would still require hundreds of whole skull CTs in order to sufficiently learn its underlying data generating distribution. The second challenge necessitates access to large GPU clusters. This challenge is perhaps the most significant bottleneck due to the high cost and scarcity of powerful GPUs required for training, though incorporating physics-informed loss functions could potentially reduce this cost.

Often transcranial clinical procedures are paired with MR scans for focal spot verification. In an attempt to further improve clinical efficiency and safety, several models have been proposed to simulate CT scans using MR scans \cite{Miscouridou2022ClassicalSimulation, Boulanger2021DeepReview, Kim2024ClinicalMRCAT}. Simulating CT from MR has the advantage of sparing the patients the ionizing radiation accompanying CT. Following the existing literature, one could envision building a conditional 3D variant of SkullGAN, where the input is the 3D MR scan and the output is its corresponding 3D CT scan. Subsequently, a 3D version of TUSNet, perhaps using vision transformers \cite{Dosovitskiy2020AnScale} instead of LSTM-Conv cells, would take as input the simulated CT of the clinically obtained MR. Though we tailored LSTM-Conv cells to the sequential nature of ultrasound propagation, vision transformers may improve results by reducing the inductive bias of convolutional layers and better capturing long-range dependencies in 3D. Alternatively, one could also design a variant of TUSNet that entirely bypasses the need for CT (either real or synthetic) and instead takes as input ZTE (zero echo time) or UTE (ultra-short echo time) MR scans \cite{Wiesinger2022Zero-TENeck, Larson2016Ultrashort7T, Kaufman2024UseImaging}, performing phase aberration correction on MR scans of the skull rather than CT.  

Due to the inherent complexity of the problem, different research groups have held different sets of variables fixed and instead focused on a narrow part of the problem. While Choi et al. have developed a novel model that outputs single-element transducer orientation and location given the desired target pressure profile with a reasonable accuracy, it does not perform phase aberration correction, does not replicate the pressure field, and the choice of target demarcation is limited to the three skulls in their physics-based simulation repository \cite{Choi2022DeepConcept}. Shin et al. developed a robust and versatile model that takes low resolution physics-based pressure field simulations as input and outputs higher resolution pressure fields. Although fairly generalizable, their proposed model does not perform end-to-end phase aberration correction or pressure field estimation and heavily relies on first running a low resolution simulation in order for the model to convert to higher resolution \cite{Shin2023Multivariable-IncorporatingSimulation}. This approach still ties their proposed model to simulations and their associated long run-times in the event of incorporating phase aberration correction. In training TUSNet, we emphasized the generalizability of TUSNet to different skulls in calculating phase aberration correction as well as the absolute pressure fields, and instead restricted our targets to only 56 locations per skull. In the Appendix, we have provided a summary comparison between these models. 
A natural next step would be to train over random targets, rather than 56 fixed targets, at finer spatial resolutions. This would generalize TUSNet to different skulls as well as virtually any target location within the region of interest. 

In summary, TUSNet has demonstrated the feasibility of applying neural networks to TUS simulations, yielding fast, generalizable, and highly accurate simulations of the ultrasound pressure field and transducer phase aberration corrections. As the first end-to-end deep learning approach to this problem, TUSNet represents an early but crucial step in the path to real-time clinical simulations with very high precision.

%%==================================%%
%%              methods             %%
%%==================================%%

\backmatter
\section*{Methods}\label{sec11}

\subsection*{Model}

\begin{figure*}[h]
\centering
\includegraphics[width=1.0\textwidth]{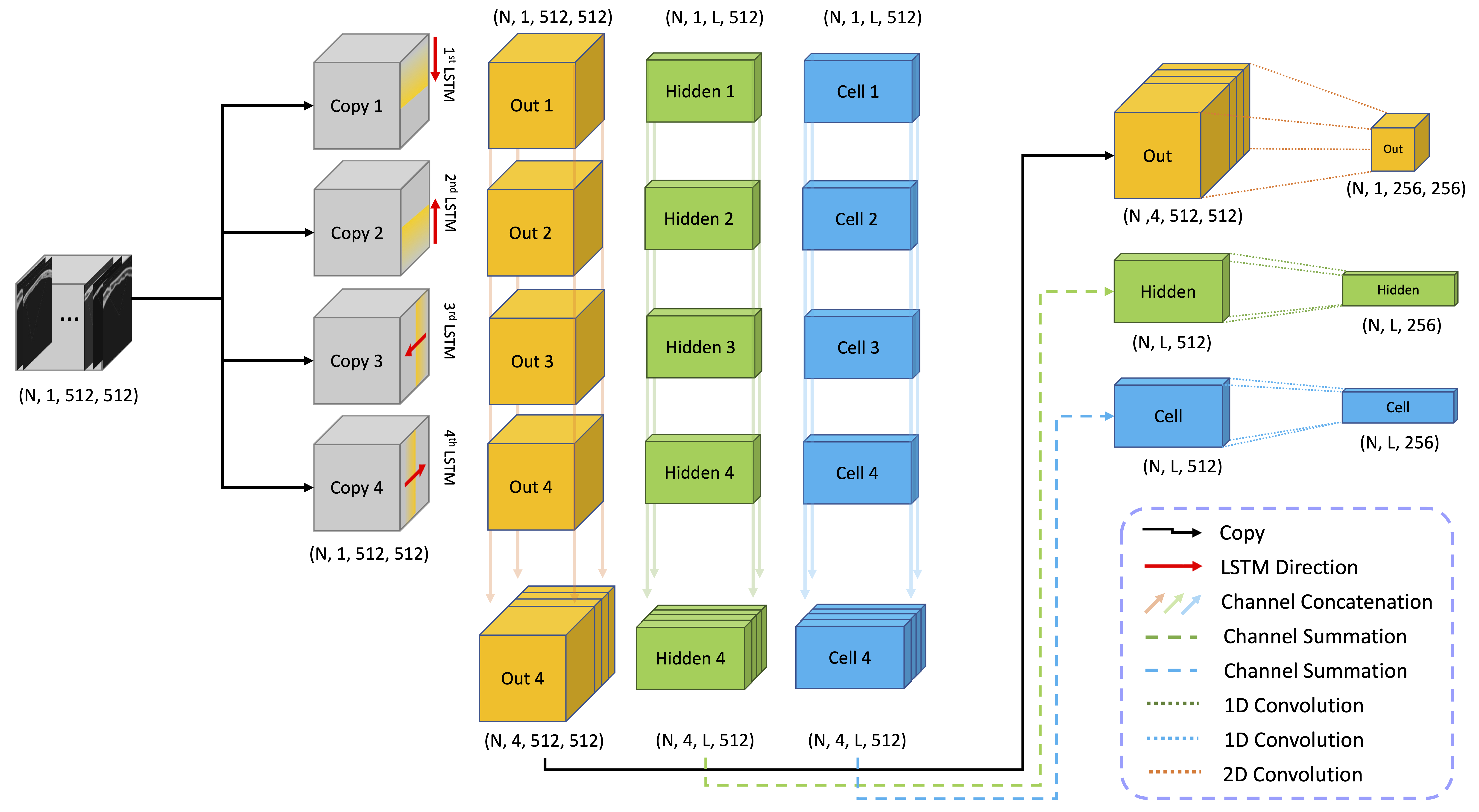}
\caption{\textbf{LSTM-Conv architecture built into the larger TUSNet model.} The LSTM-Conv cell unit consists of four long short-term memory (LSTM) cells that scan the input images in four directions, as indicated by the red arrows. The outputs of this step are concatenated along their channel dimension and passed through further convolutional layers to reduce the dimensionality of the output. Every LSTM-Conv cell produces an output, a hidden state, and a cell state, all of which will be used as input to the subsequent LSTM-Conv cell in a series of five such cells in the encoding block, and five such cells in the decoding block of TUSNet.}\label{LSTM-Conv}
\end{figure*}

At the heart of TUSNet lies the LSTM-Conv cell (detailed in Fig. \ref{LSTM-Conv}), fusing the strengths of Long Short-Term Memory (LSTM) networks with Convolutional Neural Networks (CNN). The LSTM layer captures the spatial, sequential order inherent to ultrasound pressure fields, while the convolutional layers condense this data into compact representations. Each LSTM-Conv cell initiates with four LSTM units, each scanning the input from a different direction (i.e., top-bottom, left-right, and their reverses). These LSTM units have equal hidden sizes that match the input size, and four layers in the final model.

The outputs of these four LSTM units are then concatenated and funneled through two convolutional layers, which alter the dimensionality of the embedding. The first convolutional layer has eight channels and maintains the dimensionality of the input, while the second one expands this to sixteen channels while either downsampling or upsampling (corresponding to cells in the encoder and decoder, respectively) the input size by a factor of two.

Finally, a pooling convolutional layer with a $1\times 1$ filter reduces the channel size back to one, ensuring the output matches the required dimensions for the next LSTM-Conv cell in the model. To enhance the network's stability and learning capability, batch normalization and a Rectified Linear Unit (ReLU) activation function are applied following each convolutional layer, including the final output of the cell. Additionally, a dropout technique with a rate of 0.2 is used to augment the network's robustness and curtail overfitting.

\begin{figure*}[h!]
\centering
\includegraphics[width=0.99\textwidth]{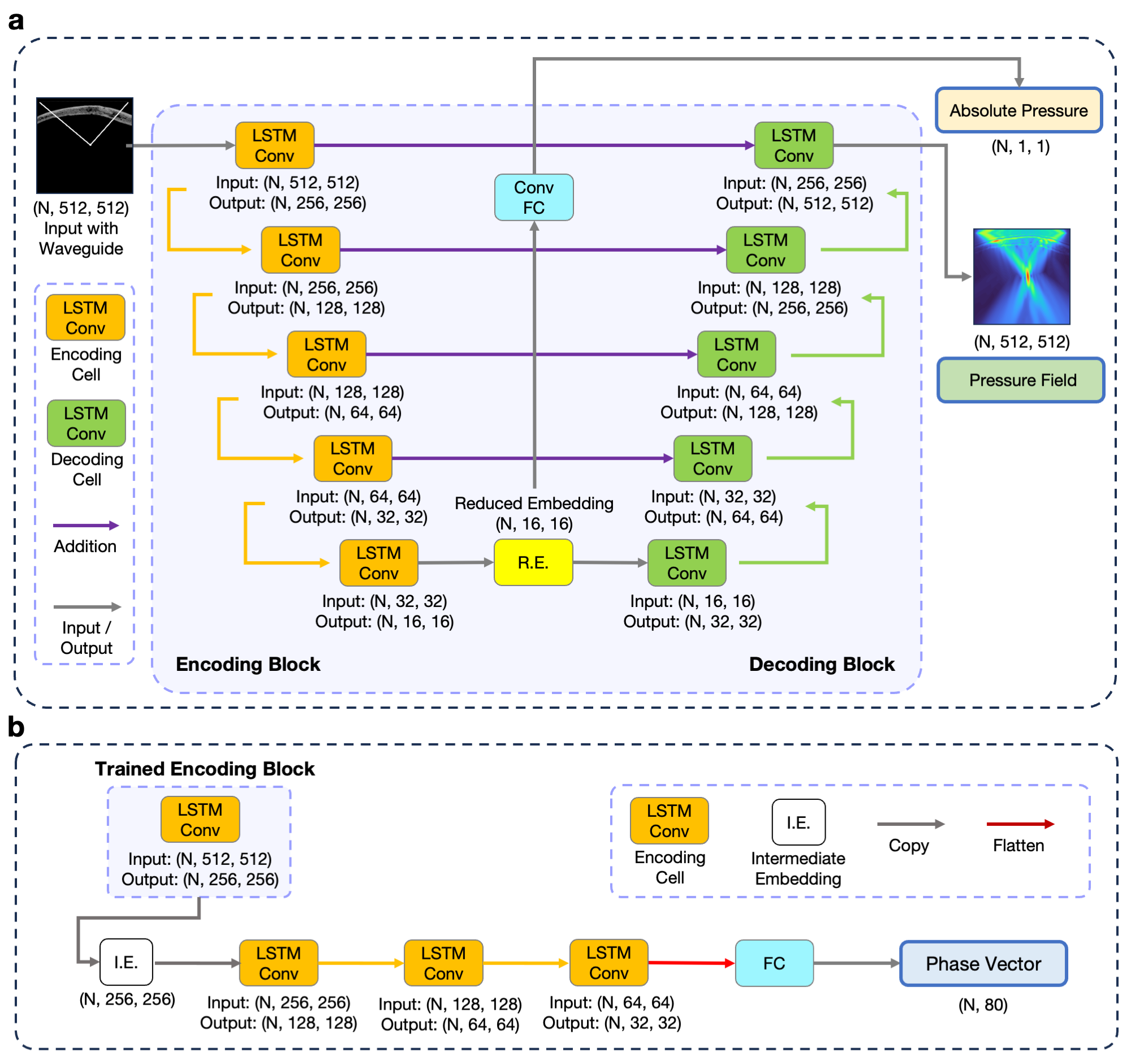}
\caption{\textbf{TUSNet architecture.} \textbf{a.} TUSNet consists of a multi-task encoder-decoder layout, along with skip connections between the two halves of the network. The output of the decoding block is the normalized Pressure Field corresponding to the input skull CT segment and target location. The reduced embedding, R.E., after passing through a series of convolutional and fully connected layers, produces the Absolute Pressure. This scalar value is used to revert back from the normal Pressure Field to the absolute pressure field in Pascals. \textbf{b} The output of the first LSTM-Conv cell of the trained encoding block is copied to an intermediate embeddings (I.E.). Subsequent series of LSTM-Conv cells are applied, followed by fully connected layers, to yield a vector of length 80, representing the phase corrections for each of the 80 transducer 
elements. Note that the weights of the LSTM-Conv cell borrowed from the trained encoding block are frozen and not updated in this process.} 
\label{tusnet_architecture}
\end{figure*}

TUSNet is structured as a symmetrical series of LSTM-Conv cells, as shown in Figure \ref{tusnet_architecture}a. It is composed of an encoding and a decoding block, each containing five LSTM-Conv cells, thereby making a total of ten cells. Starting with the input, the encoding block gradually downsamples the data, transforming and compressing it into a more compact representation. Each LSTM-Conv cell in this block takes an input size that halves at each successive layer, aligning with the notion of downsampling. Similarly, the decoding block consists of LSTM-Conv cells which receive a progressively larger input size, yielding an upsampling process that attempts to rebuild the original data from the condensed representation (i.e. the Reduced Embedding in Fig. \ref{tusnet_architecture}a).

TUSNet also incorporates skip-connections \cite{He2015DeepRecognition}, which link corresponding layers from the encoder and decoder blocks, allowing direct information transfer between these layers. This method is employed to circumvent the issue of information loss due to vanishing gradients during the dense encoding and decoding processes. This approach allows TUSNet to effectively learn compact representations of the skull CT that can be used to efficiently recover the phase aberration corrections required for the transducer, while retaining the level of information required to reconstruct highly detailed and accurate pressure fields.  

The phase decoder, described in Figure \ref{tusnet_architecture}b, uses the trained TUSNet encoding block to reconstruct the optimal phase aberration corrections for a given skull. Initiating this process, the phase decoder passes the output of the first LSTM-Conv cell of the trained encoding block through three additional LSTM-Conv layers. This operation downscales the embedding to a lower-dimensional feature space, while still allowing the model to use more information from the input. 
After flattening the output of the last LSTM-Conv cell, a final series of fully-connected layers are applied to produce the predicted phase delays. 

A simpler decoder, using only the reduced embedding produced by the encoder, is used to predict the peak pressure. Given the simplicity of having to estimate only a scalar value (Absolute Pressure in Figure \ref{tusnet_architecture}a), a series of convolutional and fully connected layers applied to the reduced embedding were sufficient in estimating the Absolute Pressure output of TUSNet.

\subsection*{Ground Truth Simulations}
All ground truth simulations were generated using the MATLAB acoustic simulation package k-Wave \cite{Treeby2010K-Wave:Fields}, a widely used tool in the field. The simulation window was set to $6\ \mathrm{cm}\times 6\ \mathrm{cm}$, or $512 \times 512$ pixels, yielding a grid size of $0.117\ \mathrm{mm}$. A flat, 80-element phased-array transducer with $0.7\ \mathrm{mm}$ spacing between its elements and total width of $5.5\ \mathrm{cm}$ was simulated. The transducer was operated at 500 kHz with a pressure of 1 MPa per transducer element, and set to output a 5-cycle ultrasound pulse over $10\ \mu s$. 

Acoustic properties were determined using the built-in k-Wave function \texttt{hounsfield2density}, which applies a linear mapping from Hounsfield Units (HU) to density. The acoustic velocity was calculated using a linear mapping of $c = 1.33\rho + 166.67$ \cite{Marquet2009Non-invasiveResults}. The density was clipped at a lower bound of 997 kg/m$^3$ to ensure that areas outside of the skull reflected the density of water. A constant attenuation of 13.3 \nicefrac{dB}{MHz$\cdot$cm} was applied to the masked skull \cite{Pinton2011AttenuationBone, Miscouridou2022ClassicalSimulation}.

Every input to k-Wave had three components: the 2D skull CT segment, the location of the transducer elements, and the target location. A skull CT slice, represented as a $512\times 512$ array, served as the primary canvas. These skull CT slices were entirely synthetic. In order to train TUSNet, we generated 5,000 \emph{synthetic} skull CT slices using SkullGAN, a generative adversarial network that outputs realistic $128\times 128$ synthetic 2D skull CT segments \cite{Naftchi-Ardebili2023SkullGAN:Networks}. Along with allowing the generation of substantial datasets, much larger than those obtainable from real datasets alone, this approach mitigates concerns of privacy as no real patient data was used to train TUSNet. 
The transducer elements were located $3\ \mathrm{mm}$ from the upper row of this canvas. The location of the target was denoted by a single pixel along with two straight lines, connecting the two outermost transducer elements to this single pixel (waveguides, see Figure \ref{tusnet_outputs}a). These waveguides, computed using the anti-aliased Bresenham's line algorithm \cite{Zingl2012ACurves}, depicted the unaberrated propagation paths for the ultrasound waves and helped improve the quality of the simulations. As we examine in the appendix through an ablation study, these waveguides serve to enhance the network's ability and efficiency in identifying the ultrasound propagation path and consequently, the target location.

% , with its value set to the global maximum HU present in the entire skull dataset. 

Next, ground truth phase aberration corrections were computed using time reversal \cite{Fink2000Time-reversedAcoustics}, wherein an initial simulation was run by sending a test pulse from the intended target to the transducer, and recording the receive delay at each element. These phase delays were then applied to the transducer, and the simulation was run forward to produce the steady-state phase-corrected pressure field. This sequence had a computation time of about 25.8 seconds per simulation on a NVIDIA A4000 GPU.

These k-Wave simulations were further processed before being used for training TUSNet: All inputs to TUSNet were normalized to the maximum HU intensity present in the entire dataset of skull CTs. Every output pressure field was normalized to itself, such that they all took values between 0 and 1. These normalized pressure fields, together with their normalization factors (respective peak values), were presented as decoupled targets to TUSNet, so that the network could learn to output both the normalized pressure field as well as its corresponding peak value of the pressure. This approach allowed for reconstructing the absolute pressure in Pascals, rather than mere normalized values that bear no physical or physiological importance.

\subsection*{Training}

After generating the 5,000 synthetic slices, to avoid overfitting to a particular skull shape or skewing the training set, the mean squared error (MSE) between each of the possible synthetic pairs was used to filter out similar skull segments and encourage high-diversity among
the slices, which resulted in 3,222, $128\times 128$ unique skull segments. After upsampling each synthetic CT slice to $512\times 512$ (using nearest-neighbor interpolation), we simulated 56 different target locations for each CT slice. This resulted in a total of $3,222\times 56=180,432$ \emph{synthetic} skull CT simulations to train TUSNet. The training paradigm was supervised, where the input comprised only the skull CT segment, transducer elements, and target location. The corresponding outputs (labels) consisted of normalized phase-corrected pressure field, absolute pressure scaling factor, and vector of phase corrections.

For testing purposes, similar simulations were run on a dataset of 22 \emph{real} skull CT slices, taken from 3 completely separate patients, to produce a test set consisting of $22\times 56=1,232$ samples. CT slices exhibiting imaging artifacts or other errors, that accounted for roughly 50\% of the data, were removed when generating the test set. It should be emphasized that this test set was never seen either by SkullGAN or by TUSNet. 

TUSNet's training was structured in a sequential manner, aiming to separately minimize the loss for each task. First, the encoder/decoder architecture was trained to generate normalized ultrasound pressure fields (\emph{Pressure Field} in Figure \ref{tusnet_architecture}a). Subsequently, as shown in Figure \ref{tusnet_architecture}a and \ref{tusnet_architecture}b, the embeddings produced by the trained encoder were used to train the other arms, to decode the phase corrections (\emph{Phase Vector}) and simulated peak absolute pressure (\emph{Absolute Pressure}). Because we did not use a physics-informed loss function, these three tasks were effectively decoupled from one another and the only shared parcel of information among them was contained in the encoding block.

All training was performed on an accelerator-optimized Google Compute Platform instance with 4 $\times$ NVIDIA A100 80GB GPUs. The model was trained using the Adam optimizer with an initial learning rate of 2e-4. The learning rate was scheduled to decrease by a factor of 0.1 when the validation loss ceased to improve over a patience of 2 epochs, with a threshold of 1e-3.

Transfer learning and fine-tuning were also employed to improve the performance of the final model: Initially, TUSNet was trained to convergence with 1 layer in each LSTM. Upon the completion of this preliminary training, the weights from this single-layered model were transferred to initialize and fine-tune a more complex 4-layer LSTM model, which was again trained to convergence. Training was carried out for approximately 50 epochs with a batch size of 256, using a weighted mean squared error (MSE) loss function for the main encoder/decoder, and L1 loss for the phase and absolute pressure decoders. The weighting aimed to enhance the focus's significance, ensuring the model prioritized achieving accurate focus in terms of pressure and location. This approach allowed the model to allocate more resources to on-target accuracy, even if it meant sacrificing some precision in estimating off-target pressures, which are less critical due to their negligible energy deposition. Early stopping was also employed as needed to prevent overfitting.  

\subsection*{TUSNet Output}
The \emph{Pressure Field}, $512\times 512$, is the normalized phase corrected field, solely tasked with providing a visual to the target shape, location, and wave distortions. The \emph{Absolute Pressure}, on the other hand, is the scalar multiplier that converts these normalized pressure values to their absolute values, making it possible to retrieve the actual pressure at the target in Pascals. Together, they provide a complete picture: \textbf{Absolute Pressure Field} $=$ \emph{Pressure Field} $\times$ \emph{Absolute Pressure} (Fig. \ref{intro}c).

The \emph{Phase Vector}, is the list of the 80 time delays that should be applied to each of the 80 transducer elements in order to correct for phase aberrations caused by the skull. To avoid confusion, we refer to this phase vector output as the \textbf{Phase Aberration Correction} of TUSNet. In our analysis, we used k-Wave for transcranial simulations, and compared the pressure field simulated using the \emph{Phase Vector} delays to the ground truth simulated with time reversal \cite{Fink2000Time-reversedAcoustics} (Fig. \ref{intro}c). 

\subsection*{Evaluation Metrics}

A variety of metrics were used to assess the performance of TUSNet, calculated for both the TUSNet pressure field and the k-Wave simulations of the pressure field with TUSNet's phase aberration corrections. The latter served to evaluate the accuracy of phase aberration correction performance of TUSNet, relying on k-Wave as a proxy for the real transducer (Fig. \ref{intro}c).

The evaluations involving the focal area and focal overlap entailed generating binary masks of ellipses fit to the pressure fields at full width at half maximum (FWHM) of the pressure, using Python's OpenCV package. 

Let $P$ be the pressure field and $P_{max}$ be the peak pressure. This ellipse-shaped binary mask, $M$, satisfied the following:

\[
M \in \{0, 1\}^{512\times 512}
\]
\[
\begin{cases} 
      M_{i,j} = 1 & \text{if } P_{i,j} \geq 0.5 \times P_{\text{max}} \\
      M_{i,j} = 0 & \text{otherwise}
\end{cases}
\]
\linebreak

%\newline

\noindent \textbf{Focal Area Error}: To assess the accuracy of the predicted focus area, we computed the absolute difference in the area between the predicted ($M_{\text{pred}}$) and ground truth ($M_{\text{gt}}$) masks, normalized by the ground truth area:

\[
\text{Focal Area Error} = \frac{|M_{\text{pred}} - M_{\text{gt}}|}{|M_{\text{gt}}|}\times 100.
\]

\noindent\textbf{Focal Overlap}: The Intersection over Union (IoU) quantified the overlap between the predicted ($M_{\text{pred}}$) and ground truth ($M_{\text{gt}}$) focal spots. It was given by:

\[
\text{IoU} = \frac{M_{\text{pred}} \cap M_{\text{gt}}}{M_{\text{pred}} \cup M_{\text{gt}}}\times 100.
\] \\

\noindent \textbf{Peak Pressure Error}: The error between the peak pressure predicted by TUSNet ($P_{pred}^p$) and the ground truth ($P_{gt}^p$) was calculated as:

\[
\text{Peak Pressure Error} = \frac{|P_{\text{pred}}^p - P_{\text{gt}}^p|}{P_{\text{gt}}^p} \times 100.
\]

It must be noted that this peak pressure need not be the pressure at the focus. The peak pressure is taken as the maximum pressure value in the entire simulation window. It may or may not be the same as the pressure at the focus of the transducer (Figure \ref{tusnet_metrics_pressure}).

\noindent \textbf{Focal Pressure Error}: The percent error between the focal pressure predicted by TUSNet ($P_{\text{pred}}^f$) and the ground truth ($P_{\text{gt}}^f$) was calculated using the percent error formula:

\[
\text{Focal Pressure Error} = \frac{|P_{\text{pred}}^f - P_{\text{gt}}^f|}{P_{\text{gt}}^f} \times 100.
\]

Unlike the peak pressure error, the focal pressure error evaluated the difference in pressure at the location of the ground truth pressure field's focus.

\noindent\textbf{Focal Position Error -- Euclidean Distance}: Given the coordinates of the predicted ($f_{\text{pred}}$) and ground truth ($f_{\text{gt}}$) foci, we computed their Euclidean distance in $mm$ as follows:

\begin{align*}
    \text{Euclidean Distance} &= \bigg[
\big(f(x)_{\text{pred}} - f(x)_{\text{gt}}\big)^2 +\\ & \big(f(y)_{\text{pred}}- f(y)_{\text{gt}} \big)^2 \bigg]^{\nicefrac{1}{2}}
\end{align*}

\noindent\textbf{Focal Position Error -- Modified Hausdorff Distance}: This metric gauged not only the accuracy in pinpointing the focus, but also the similarity in shape and orientation of the prediction and ground truth FWHM ellipses. The Modified Hausdorff Distance (MHD) \cite{DubuissonAMatching} enjoys other advantages such as robustness to noise and applicability to focal spots of different sizes. Let $B_{\text{pred}}$ and $B_{\text{gt}}$ be the boundaries of the largest connected components in the predicted and ground truth pressure fields, respectively. The MHD was then calculated as:

\begin{align*}
\text{MHD} = \max\Bigg\{&\frac{1}{|B_{\text{pred}}|} \sum_{b \in B_{\text{pred}}} \min_{g \in B_{\text{gt}}} d(b,g), \\
&\frac{1}{|B_{\text{gt}}|} \sum_{g \in B_{\text{gt}}} \min_{b \in B_{\text{pred}}} d(g,b)\Bigg\}.
\end{align*}

\noindent where $d(\cdot, \cdot)$ denoted the Euclidean distance. \linebreak

\noindent\textbf{Axial Focal Position Error}: Using the major and minor axes of the FWHM ellipses, we were able to calculate the tilt angle of the focal spots ($\theta$). $\theta$ to the right of the normal was considered negative, and $\theta$ to the left of the normal was considered positive. This allowed us to project the Euclidean distance in the original Cartesian coordinate system ($d_x$ and $d_y$) to the major and minor axes of the FWHM ellipse of the predicted field. Specifically, projecting onto the major axis of the predicted FWHM ellipse resulted in the axial focal position error:

\begin{align*}
\text{Axial Error} = d_y\cos(\theta) - d_x\sin(\theta).
\end{align*}

\noindent\textbf{Lateral Focal Position Error}: Similarly, the lateral component of the focal position error, as projected onto the minor axis of the FWHM ellipse of the predicted field, was obtained using:

\begin{align*}
\text{Lateral Error} = d_y\sin(\theta) + d_x\cos(\theta).
\end{align*}

\backmatter

%%==================================%%
%%          supplementary           %%
%%==================================%%

\newpage \onecolumn

%%==================================%%
%%           back matter            %%
%%==================================%%

\bmhead{Data availability}

The training set used in this paper is made available online. 

\bmhead{Code availability}

TUSNet was written in Python v3.8 using the PyTorch library. All of the source code and training data are available at https://github.com/kbp-lab/TUSNet. 

\bmhead{Acknowledgments}

We would like to thank Dr. Fanrui Fu for providing the anonymized human skull CTs we used for testing TUSNet. We would also like to thank Dr. Reza Pourabolghasem for his invaluable feedback on the manuscript.  This work was supported by NIH R01 Grant EB032743.

\bmhead{Author Contributions}

The first two authors jointly led this project and contributed equally to the design, training, and analysis of TUSNet. All authors contributed to the writing of the paper. 

% If needed...
% We have provided a full account of each author's contributions below.

% \textbf{Kasra Naftchi-Ardebili}: 

% \textbf{Karanpartap Singh}

% \textbf{Gerald R. Ppelka}

% \textbf{Kim Butts Pauly}

%%==================================%%
%%            references            %%
%%==================================%%

%\bibliography{references}

\bibliographystyle{ieeetr}

%%==================================%%
%%             appendix             %%
%%==================================%%

\begin{appendices}

% \twocolumn
\newpage
\section*{Appendix}

\subsection*{Comparison Against Other Models}
In the table below we summarize the performance of TUSNet against the other two models referenced in the paper. Direct head to head comparison is difficult, as none of these models tackle the exact same problem under identical conditions. However, a high level comparison of their fixed and variable parameters, outputs, run times, and accuracies gives a sense of the challenges that remain to be addressed. 

\begin{table*}[h]
\resizebox{\textwidth}{!}{%
\begin{tabular}{|l|l|l|l|l|l|l|l|l|l|l|l|l|}
\hline
\multicolumn{1}{|c|}{Model} &
  \multicolumn{1}{c|}{Input} &
  \multicolumn{1}{c|}{Output} &
  \multicolumn{1}{c|}{Dim.} &
  \multicolumn{1}{c|}{\begin{tabular}[c]{@{}c@{}}Pressure\\ Field\end{tabular}} &
  \multicolumn{1}{c|}{IoU} &
  \multicolumn{1}{c|}{\begin{tabular}[c]{@{}c@{}}Positioning\\ Error\end{tabular}} &
  \multicolumn{1}{c|}{\begin{tabular}[c]{@{}c@{}}Focal \\ Pressure\\ Error\end{tabular}} &
  \multicolumn{1}{c|}{\begin{tabular}[c]{@{}c@{}}Phase \\ Aberration\\ Correction\end{tabular}} &
  \multicolumn{1}{c|}{\begin{tabular}[c]{@{}c@{}}Generalizes\\ to Various \\ Skulls\end{tabular}} &
  \multicolumn{1}{c|}{\begin{tabular}[c]{@{}c@{}}Generalizes \\ to Various \\ Target\end{tabular}} &
  \multicolumn{1}{c|}{\begin{tabular}[c]{@{}c@{}}Inference\\ Run Time\end{tabular}} &
  \multicolumn{1}{c|}{\begin{tabular}[c]{@{}c@{}}Machine\\ Specifications\end{tabular}} \\ \hline
SE-SRResNet \cite{Shin2023Multivariable-IncorporatingSimulation} &
  \begin{tabular}[c]{@{}l@{}}1.0 mm transcranial \\ simulation \\ (using FDTD algorithm)\end{tabular} &
  \begin{tabular}[c]{@{}l@{}}0.5mm transcranial \\ pressure field\end{tabular} &
  3D &
  normalized &
  81\% &
  NA &
  NA &
  NA &
  Yes &
  Yes &
  3840 ms &
  \begin{tabular}[c]{@{}l@{}}one NVIDIA A100 \\ 40 GB\\ Tensor Core GPU\end{tabular} \\ \hline
CNN+AC \cite{Choi2022DeepConcept}&
  \begin{tabular}[c]{@{}l@{}}binarized focal \\ ellipsoid at the \\ desired target\end{tabular} &
  \begin{tabular}[c]{@{}l@{}}optimal location for \\ transducer placement\end{tabular} &
  3D &
  normalized &
  74.49\% &
  0.96 mm &
  NA &
  NA &
  No &
  No &
  12.25 ms &
  \begin{tabular}[c]{@{}l@{}}Intel i9-7940X-CPU, \\ 64.0 GB-RAM, and \\ NVIDIA GeForce \\ 2080Ti-single GPU\end{tabular} \\ \hline
TUSNet &
  \begin{tabular}[c]{@{}l@{}}skull CT segment, \\ target location,\\ and transducer location\end{tabular} &
  \begin{tabular}[c]{@{}l@{}}phase corrected \\ transcranial \\ pressure field\end{tabular} &
  2D &
  \begin{tabular}[c]{@{}l@{}}absolute \\ (in Pascals)\end{tabular} &
  85.58\% &
  0.59 mm &
  1.71\% &
  Yes &
  Yes &
  No &
  20.7 ms &
  \begin{tabular}[c]{@{}l@{}}one NVIDIA A4000\\ 16GB GPU\end{tabular} \\ \hline
  k-Wave &
  \begin{tabular}[c]{@{}l@{}}skull CT segment, \\ target location,\\ and transducer location\end{tabular} &
  \begin{tabular}[c]{@{}l@{}}phase corrected \\ transcranial \\ pressure field\end{tabular} &
  2D &
  \begin{tabular}[c]{@{}l@{}}absolute \\ (in Pascals)\end{tabular} &
  \begin{tabular}[c]{@{}l@{}}ground \\ truth\end{tabular} &
  \begin{tabular}[c]{@{}l@{}}ground \\ truth\end{tabular} &
  \begin{tabular}[c]{@{}l@{}}ground \\ truth\end{tabular} &
  Yes &
  Yes &
  Yes &
  25800 ms &
  \begin{tabular}[c]{@{}l@{}}one NVIDIA A4000\\ 16GB GPU\end{tabular} \\ \hline
\end{tabular}}
\caption{\textbf{TUSNet performance comparison against SE-SRResNet, CNN+AC, and k-Wave (which was used as the ground truth reference in training TUSNet).} The current proof-of-concept version of TUSNet is a 2D model, whereas the other two models are in 3D. As such, it is hard to directly compare its superior performance in IoU and positioning error to the other two models. Focal pressure error cannot be compared at all, as the other two models report their final pressure fields in normalized scale. Similarly, the inference run times cannot be compared directly, as the machines on which these models were tested were quite different. However, on a head-to-head comparison with k-Wave, on the same hardware, TUSNet inference time is over 1200 times faster.}
\label{tusnet_metrics_table_1}
\end{table*}

\subsection*{Ablation Study}
We made several claims on the superiority of our model architecture to conventional approaches, such as upweighting the loss around the focus compared to other areas of the field of view (weighted MSE) and using waveguides rather than point targets to denote the intended target. In the below table, we compare the performance of TUSNet's absolute pressure field output to its modified variants:
\begin{itemize}
    \item MSE Loss: under this ablation category, we replaced the weighted MSE with a regular MSE. 
    \item Point Target: under this category, we removed the waveguides from the input and represented the target with only a single point.  
\end{itemize}

The full TUSNet outperformed its modified variants across almost all evaluation metrics. If we were to rank the significance of these different components based on their performance degradation, we would conclude that inclusion of waveguides is the most crucial component, followed by the adoption of a weighted MSE instead of regular MSE loss. 

\begin{table*}[h]
\resizebox{\textwidth}{!}{%
\begin{tabular}{|c|cccccccc|}
\hline
\multirow{2}{*}{} &
  \multicolumn{8}{c|}{Absolute Pressure Field Performance on the Test Set} \\ \hhline{|~|--------|}    
 &
  \multicolumn{2}{c|}{Focal Area Error (\%)} &
  \multicolumn{2}{c|}{Pressure Error (\%)} &
  \multicolumn{4}{c|}{Focal Position Error (mm)} \\ \hline
Ablation: &
  \multicolumn{1}{c|}{Percent Error} &
  \multicolumn{1}{c|}{IoU ($\uparrow$)} &
  \multicolumn{1}{c|}{Focal} &
  \multicolumn{1}{c|}{Peak} &
  \multicolumn{1}{c|}{Euclidean} &
  \multicolumn{1}{c|}{Hausdorff} &
  \multicolumn{1}{c|}{Axial} &
  Lateral \\ \hline
MSE Loss &
  \multicolumn{1}{c|}{10.47 $\pm$ 6.22} &
  \multicolumn{1}{c|}{84.90 $\pm$ 7.18} &
  \multicolumn{1}{c|}{-} &
  \multicolumn{1}{c|}{-} &
  \multicolumn{1}{c|}{0.43 $\pm$ 0.34} &
  \multicolumn{1}{c|}{0.19 $\pm$ 0.08} &
  \multicolumn{1}{c|}{0.42 $\pm$ 0.35} &
  0.06 $\pm$ 0.08 \\ \hline
Point Target &
  \multicolumn{1}{c|}{17.92 $\pm$ 16.23} &
  \multicolumn{1}{c|}{77.04 $\pm$ 14.53} &
  \multicolumn{1}{c|}{-} &
  \multicolumn{1}{c|}{-} &
  \multicolumn{1}{c|}{0.35 $\pm$ 0.26} &
  \multicolumn{1}{c|}{0.30 $\pm$ 0.20} &
  \multicolumn{1}{c|}{0.33 $\pm$ 0.27} &
  0.06 $\pm$ 0.07 \\ \Xhline{5\arrayrulewidth}
Full TUSNet &
  \multicolumn{1}{c|}{\textbf{8.27 $\pm$ 5.43}} &
  \multicolumn{1}{c|}{\textbf{86.95 $\pm$ 7.05}} &
  \multicolumn{1}{c|}{\textbf{6.07 $\pm$ 4.0}} &
  \multicolumn{1}{c|}{\textbf{5.87 $\pm$ 3.95}} &
  \multicolumn{1}{c|}{\textbf{0.30 $\pm$ 0.26}} &
  \multicolumn{1}{c|}{\textbf{0.15 $\pm$ 0.08}} &
  \multicolumn{1}{c|}{\textbf{0.28 $\pm$ 0.26}} &
  \textbf{0.06 $\pm$ 0.07} \\ \hline
\end{tabular}}
\caption{\textbf{TUSNet Absolute Pressure Field performance on the test set under various ablations on the full TUSNet.} \emph{MSE loss} denotes the scenario where we replaced the weighted MSE with a regular MSE. \emph{Point Target} refers to designating the target as a single point without the linear wave guides. \emph{Full TUSNet} denotes the best performing TUSNet model presented in this paper, without any of these ablations.}
\label{tusnet_metrics_table_1}
\end{table*}

Along with ablating components of the main encoder/decoder portion of TUSNet, tasked with predicting the pressure field, we also investigated the impact of using LSTM-Conv cells in the phase decoder in comparison to simple convolutions. To this end, we saw a significant improvement in TUSNet's phase aberration correction accuracy across all metrics when LSTM-Conv cells were employed, highlighting the benefits of applying sequence models to this task through our architecture.

\begin{table*}[h]
\resizebox{\textwidth}{!}{%
\begin{tabular}{|c|cccccccc|}
\hline
\multirow{2}{*}{} &
  \multicolumn{8}{c|}{Phase Aberration Correction Performance on the Test Set} \\ \hhline{|~|--------|}    
 &
  \multicolumn{2}{c|}{Focal Area Error (\%)} &
  \multicolumn{2}{c|}{Pressure Error (\%)} &
  \multicolumn{4}{c|}{Focal Position Error (mm)} \\ \hline
Ablation: &
  \multicolumn{1}{c|}{Percent Error} &
  \multicolumn{1}{c|}{IoU ($\uparrow$)} &
  \multicolumn{1}{c|}{Focal} &
  \multicolumn{1}{c|}{Peak} &
  \multicolumn{1}{c|}{Euclidean} &
  \multicolumn{1}{c|}{Hausdorff} &
  \multicolumn{1}{c|}{Axial} &
  Lateral \\ \hline
Conv Only &
  \multicolumn{1}{c|}{26.14 $\pm$ 16.96} &
  \multicolumn{1}{c|}{60.88 $\pm$ 17.75} &
  \multicolumn{1}{c|}{14.16 $\pm$ 14.07} &
  \multicolumn{1}{c|}{2.98 $\pm$ 2.11} &
  \multicolumn{1}{c|}{1.56 $\pm$ 1.10} &
  \multicolumn{1}{c|}{0.60 $\pm$ 0.35} &
  \multicolumn{1}{c|}{1.39 $\pm$ 1.13} &
  0.49 $\pm$ 0.43 \\ \Xhline{5\arrayrulewidth}
LSTM-Conv &
  \multicolumn{1}{c|}{\textbf{8.52 $\pm$ 4.61}} &
  \multicolumn{1}{c|}{\textbf{85.58 $\pm$ 5.85}} &
  \multicolumn{1}{c|}{\textbf{1.71 $\pm$ 1.09}} &
  \multicolumn{1}{c|}{\textbf{1.13 $\pm$ 0.68}} &
  \multicolumn{1}{c|}{\textbf{0.59 $\pm$ 0.35}} &
  \multicolumn{1}{c|}{\textbf{0.18 $\pm$ 0.07}} &
  \multicolumn{1}{c|}{\textbf{0.57 $\pm$ 0.35}} &
  \textbf{0.10 $\pm$ 0.10} \\ \hline
\end{tabular}}
\caption{\textbf{TUSNet Phase Aberration Correction performance on the test set under ablation study for the LSTM-Conv architecture.} \emph{Conv Only} denotes the scenario where we used only a network of convolutional and fully-connected layers to predict the optimal phase aberration correction. \emph{LSTM-Conv} denotes the TUSNet phase decoder presented in this paper, which employs LSTM-Conv and fully-connected layers to estimate the phase aberration correction.}
\label{tusnet_metrics_table_1}
\end{table*}

\newpage
\subsection*{Worst Cases}

While TUSNet demonstrated remarkable performance across multiple metrics, there were instances where it underperformed, particularly in cases with high anatomical complexity or poor image quality. These worst cases, while rare, provide crucial insights into the limitations and potential areas for improvement of the model. Here, we examine the worst-performing samples for each metric evaluated — focal area, pressure, and positioning errors.

In the most challenging cases, TUSNet struggled to replicate the focal area as accurately due to irregular skull geometries that were not well-represented in the training dataset, achieving an IoU around 60\% for both the absolute pressure field and phase aberration correction. A larger number of real skull CTs for training SkullGAN, thereby increasing the diversity of our synthetic skull-based training set, would help mitigate this problem by better representing diverse skull geometries.

As for positional accuracy, even the worst cases exhibited an error under 1mm. Particularly, for the phase aberration correction worst case, TUSNet's focal spot was inscribed within the ground truth focal spot. However, the Modified Hausdorff Distance reported a 0.79mm error due to the mismatch between the ellipses' shapes.

Pressure estimation presented challenges in cases with pronounced energy loss at the skull, resulting in an unexpected pressure within the brain. The worst case for the absolute pressure field had a focal pressure of 22\%, with 7.4\% for the phase aberration correction. Additional training on a more diverse population of skull CTs would likely improve these worst-cases.

\begin{figure*}[h]
\centering
\includegraphics[width=1.0\textwidth]{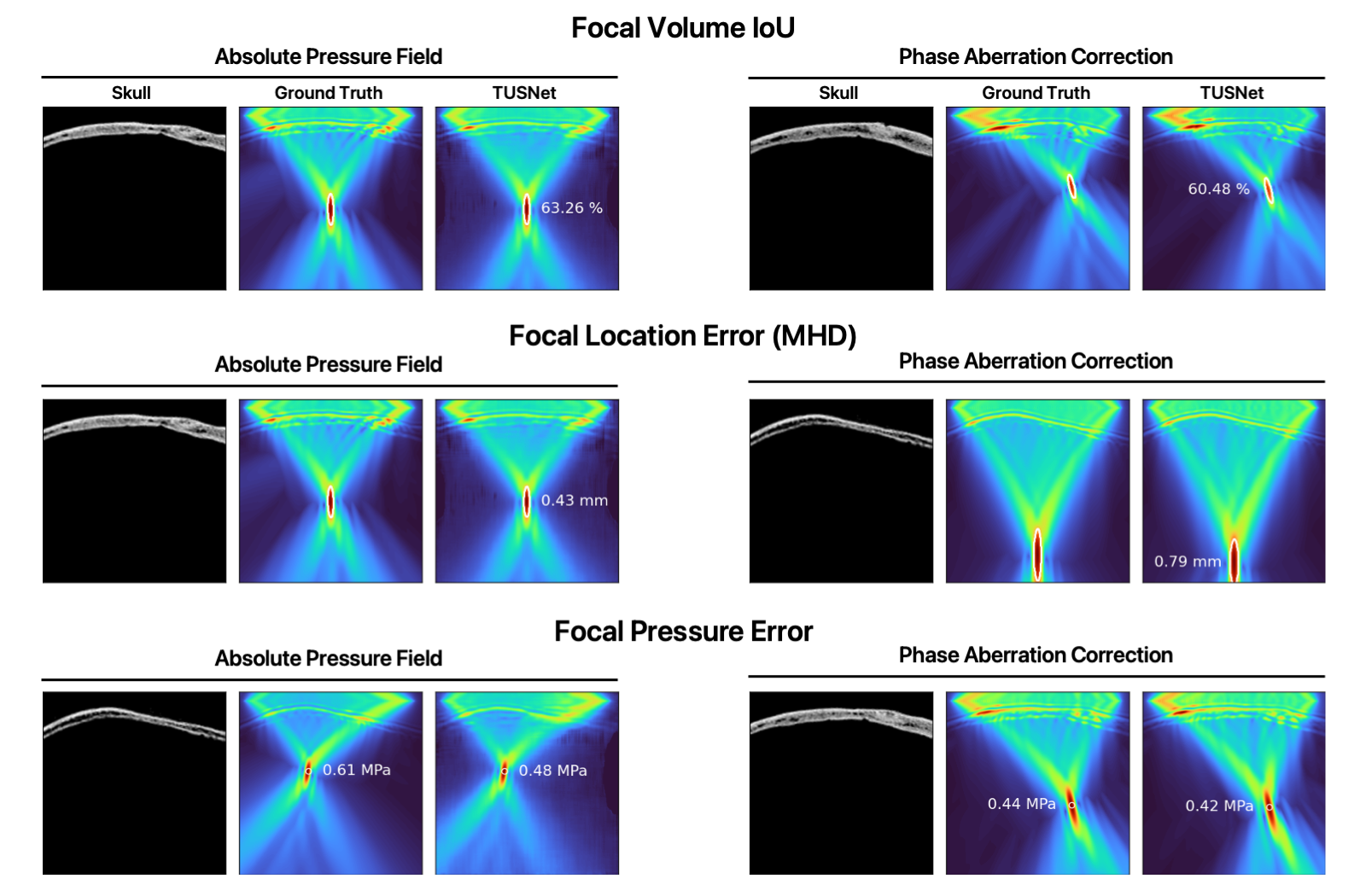}
\caption{\textbf{TUSNet worst-case performance for each of the metrics (focal area, pressure, and positioning error) examined in the paper.} The input skull, ground truth (from k-Wave), and TUSNet output are given for the worst-performing sample in both the absolute pressure field and phase aberration correction.}
\label{tusnet_worst_cases}
\end{figure*}

\end{appendices}

\end{document}